\documentclass[amssymb,twocolumn,aps]{revtex4}
\usepackage{amsmath}
\usepackage{enumerate}
\begin{document}
\def \bnab {{\bf\nabla}\hskip-8.8pt{\bf\nabla}\hskip-9.1pt{\bf\nabla}}
\title{Energy--Momentum Tensor for the Electromagnetic Field in
a Dielectric}
\author{Michael E. Crenshaw and Thomas B. Bahder}
\affiliation{RDMR-WSS, Aviation and Missile RDEC, US Army RDECOM,
Redstone Arsenal, AL 35898, USA}
\date{\today}
\begin{abstract}
The total momentum of a thermodynamically closed system is unique, as is
the total energy.
Nevertheless, there is continuing confusion concerning the correct form
of the momentum and the energy--momentum tensor for an electromagnetic
field interacting with a linear dielectric medium.
Rather than construct a total momentum from the Abraham momentum or the
Minkowski momentum, we define a thermodynamically closed system
consisting of a propagating electromagnetic field and a negligibly
reflecting dielectric and we identify the Gordon momentum as
the conserved total momentum by the fact that it is invariant in time.
In the formalism of classical continuum electrodynamics, the
Gordon momentum is therefore the unique representation of the total
momentum in terms of the macroscopic electromagnetic fields and the 
macroscopic refractive index that characterizes the material.
We also construct continuity equations for the energy and the Gordon
momentum, noting that a time variable transformation is necessary to
write the continuity equations in terms of the densities of 
conserved quantities.
Finally, we use the continuity equations and the time-coordinate
transformation to construct an array that has the properties of a
traceless, symmetric energy--momentum tensor.
\end{abstract}
\maketitle
\vskip 3.0cm
\par
\section{Introduction}
\par
The energy--momentum tensor is a concise way to represent the
conservation properties of an unimpeded flow field.
For most types of simple flows, the energy--momentum tensor is 
well-defined, with the notable exception of the electromagnetic field in
a linear dielectric material.
The Abraham--Minkoswski controversy
\cite{BIGord,BIPeie,BIKra,BIOth1,BIChiao,BIOth2,BIEinL,BINels,BIPenHaus}
for the momentum of electromagnetic fields in a dielectric began with
the derivation of the energy--momentum four-tensor by
Minkowski \cite{BIMin}.
Noting that the unsymmetrical Minkowski tensor does not support
conservation of angular momentum,
Abraham \cite{BIAbr} proposed an energy--momentum tensor that was
symmetric, but at the expense of a new phenomenological force.
In order to address this constraint and additional issues, Einstein
and Laub \cite{BIEinL}, Nelson \cite{BINels}, and others
proposed variants of the energy--momentum tensor.
\par
The crux of the Abraham--Minkowski controversy is whether the
electromagnetic momentum density in a dielectric is of the Minkowski
form
\begin{equation}
{\bf g}_M=\frac{1}{c} ({\bf D}\times{\bf B})
\label{EQa1.01}
\end{equation}
or the Abraham form
\begin{equation}
{\bf g}_A=\frac{1}{c} ({\bf E}\times{\bf H}).
\label{EQa1.02}
\end{equation}
Experimental efforts to resolve the theoretical impasse have not
been definitive.
While some experiments favor the Abraham formula, other experiments
support Minkowski's version.
Brevick's \cite{BIBrev} analysis of experiments performed by
Jones and Richards \cite{BIExp1},
Ashkin and Dziedzic \cite{BIExp2},
and others showed that the allocation of momentum
between the field and material was the determining factor in whether a
particular experimental result was described by the Abraham or Minkowski
form of electromagnetic momentum.
Following Brevick \cite{BIBrev}, the formula for the field momentum has
been shown repeatedly to be arbitrary such that any of the
formulas for the field momentum can be combined with an appropriate
momentum for the material to produce the same total
momentum \cite{BIPfeifer,BIMikura,BIcankin}.
\par
In 1973, Gordon \cite{BIGord} constructed the total momentum from a
microscopic model in which the electromagnetic field component of the
total momentum is said to be the Abraham momentum and the dielectric is
treated as a dilute collection of electric dipoles with center-of-mass
motion in the direction of propagation of the field.
Gordon \cite{BIGord} discusses the empirical and experimental
validation of the total momentum density
\begin{equation}
{\bf g}_G = \frac{n}{c} ({\bf E}\times{\bf B})
\label{EQa1.03}
\end{equation}
and shows that the density ${\bf g}_G$, integrated over
a volume containing the entire field, is invariant in time.
Gordon concludes that the Abraham momentum density represents the true
momentum density of the electromagnetic field and the Minkowski
momentum density includes a pseudomomentum.
\par
In this article, we investigate conservation of energy and momentum in
a thermodynamically closed system consisting of the macroscopic
electromagnetic field and a negligibly reflecting linear dielectric
(such as a dielectric with an anti-reflection coating).
Here, we identify the Gordon momentum \cite{BIGord} as the total
momentum by the fact that it is invariant in time and therefore 
a conserved quantity in an isolated system.
We find that it is not necessary to decompose the Gordon momentum into
the sum of a field momentum, such as the Abraham or Minkowski momentum,
and a material momentum, such as the canonical or kinetic
momentum \cite{BIChiao,BIcankin}.
Instead, we work with the total energy and total momentum and derive
continuity equations in terms of the densities of these conserved
quantities.
The continuity equations are then used to construct an array that has
the properties of a traceless, symmetric energy--momentum
tensor, but in a coordinate system with time-like coordinate $ct/n$.
\par
\section{Energy--Momentum Tensor of Noninteracting Particles}
\par
In the continuum limit, the density of any property of identical
noninteracting particles can be treated as the number density multiplied
by the amount of the property that can be attributed to each particle.
The continuity equation corresponding to a specific property,
such as mass, charge, or energy, is then obtained by substitution of
the specific property density for a placeholder number density.
For an infinitesimal element of volume in an inviscid
sourceless flow, the continuity equation
\begin{equation}
\frac{\partial s}{\partial t}
+\nabla\cdot s{\bf u}
=0
\label{EQa2.01}
\end{equation}
is derived by applying the divergence theorem to a Taylor series
expansion of the property density field $ s$ and the vector
velocity field ${\bf u}=(u^x,u^y,u^z)$ of the flow \cite{BIFoxMcD}.
The continuity equation reflects the conservation of a continuous scalar
property in a flow in terms of the equality of the net rate of flux
out of the volume and the time rate of change of the property
density $ s$ inside the volume.
Depending on the context, the three-vector
\begin{equation}
{\bf g}= s{\bf u}
\label{EQa2.02}
\end{equation}
is known as the momentum density, the flux, or the current density of
the property.
Specifically, ${\bf g}$ corresponds to the linear momentum density if
$ s$ is a mass density and corresponds to the charge current density
if $ s$ is the electric charge density.
\par
Some conserved properties, such as momentum, are vectors. For a flow,
the density of a conserved vector property can be represented 
as ${\bf  s}=(s^x, s^y, s^z)$.
Applying the scalar formalism to the three orthogonal components of the
property density vector yields scalar continuity equations
\begin{subequations}
\label{EQa2.03}
\begin{equation}
\frac{\partial s^x}{\partial t}
+\nabla\cdot s^x{\bf u}
=0
\label{EQa2.03a}
\end{equation}
\begin{equation}
\frac{\partial s^y}{\partial t}
+\nabla\cdot s^y{\bf u}
=0
\label{EQa2.03b}
\end{equation}
\begin{equation}
\frac{\partial s^z}{\partial t}
+\nabla\cdot s^z{\bf u}
=0.
\label{EQa2.03c}
\end{equation}
\end{subequations}
\par
At this point, we adopt a four-dimensional notation where repeated
indices are summed.
We take Roman indices to run from 1 to 3 and we identify the coordinates
$x^i$ with the Cartesian coordinates, such that $x^1=x$, $x^2=y$, 
and $x^3=z$.
Greek indices run from 0 to 3 and $x^0$ is identified with the
time-like coordinate $ct$.
The Minkowski space-time metric is $diag(-1,1,1,1).$
Finally, partial differentiation with respect to the indexed coordinates
is represented by $\partial_{\alpha}=\partial/(\partial x^{\alpha})$.
\par
The four continuity equations (\ref{EQa2.01}) and (\ref{EQa2.03}) can
be concisely represented by
\begin{equation}
\partial_{\beta} G^{\alpha\beta}=0,
\label{EQa2.04}
\end{equation}
where
\begin{equation}
G^{\alpha\beta}=
\left [
\begin{matrix}
c^2 s   &cu^x s     &cu^y s     &cu^z s
\cr
c s^x   &u^x s^x   &u^y s^x   &u^z s^x
\cr
c s^y   &u^x s^y   &u^y s^y   &u^z s^y
\cr
c s^z   &u^x s^z   &u^y s^z   &u^z s^z
\cr
\end{matrix}
\right ].
\label{EQa2.05}
\end{equation}
Consider a transformation to a new set of coordinates
${x^i}^{\prime}=f_i(x^0,x^1,x^2,x^3)$.
In order for this matrix to transform as a tensor, $s^x$, $s^y$, and
$s^z$ must be expressible in terms of $s$ and components of ${\bf u}$.
For a closed system, conservation of angular momentum requires
$G^{\alpha\beta}$ to be symmetric \cite{BILL}
\begin{equation}
G^{\alpha\beta}=G^{\beta\alpha}.
\label{EQa2.06}
\end{equation}
Therefore, the vector property ${\bf s}$ must represent the flux of the
conserved scalar quantity $s$ such that ${\bf s} = s{\bf u}$.
Conversely, the flux $s{\bf u}$ of a scalar property of the particles
is a conserved vector property of the flow.
The covariant form of the continuity equation is the four-divergence,
Eq.\ (\ref{EQa2.04}), of the energy--momentum four-tensor 
\begin{equation}
G^{\alpha\beta}=
\left [
\begin{matrix}
c^2 s   &c s u^x    &c s u^y   &c s u^z
\cr
c s u^x & s u^x u^x  & s u^x u^y  & s u^x u^z
\cr
c s u^y & s u^y u^x  & s u^y u^y  & s u^y u^z
\cr
c s u^z & s u^z u^x  & s u^z u^y  & s u^z u^z
\cr
\end{matrix}
\right ].
\label{EQa2.07}
\end{equation}
The continuity equation (\ref{EQa2.04}) with the four-tensor
(\ref{EQa2.07}) is valid for any conserved extensive quantity in a
simple flow, not just for the mass of a fluid.
\par
The energy--momentum tensor given in Eq.\ (\ref{EQa2.07}) has some
essential properties.
First, the four-divergence of each row vector,
Eq.\ (\ref{EQa2.04}),
is a continuity law corresponding to the conservation of the property
represented by the property density in the first element of that row.
Second, the tensor is symmetric,
corresponding to the absence of unbalanced shear forces and conservation
of angular momentum in a closed system \cite{BILL}.
Third, as a consequence of diagonal symmetry,
\begin{equation}
\partial_{\alpha}G^{\alpha\beta}=0,
\label{EQa2.10}
\end{equation}
the four-divergence of each column vector is a continuity law
corresponding to the conservation of the property represented by the
property density in the first element of that column.
It should also be noted that this simple energy--momentum tensor is
based on the properties of an unimpeded flow.
If the flow is redirected by impact with a macroscopic object then one
is obligated to include the equations of motion of the object or
the forces of restraint.
\par
\section{The Abraham and Minkowski Energy--Momentum Tensors}
\par
The Abraham and Minkowski energy--momentum tensors are examples of a
number of different tensors that have been proposed for the
electromagnetic field in a dielectric \cite{BIPfeifer}.
The Minkowski tensor can be constructed from the continuity equations
for energy flux and momentum flux by the same procedure that was used
to construct the array (\ref{EQa2.05}).
The Minkowski tensor is not symmetric and therefore
violates angular momentum conservation if it is
the total energy--momentum tensor of a closed system.
Consequently, the Minkowski tensor is considered to be a
representation of the energy and momentum of a component
of the system \cite{BIPfeifer}.
In this section, we outline the construction of the Minkowski tensor
and describe how the procedure is modified to obtain the Abraham tensor.
Neither the Minkowski tensor, nor the Abraham tensor, satisfy the
requirements of a total energy--momentum tensor.
The Minkowski tensor is not symmetric while the Abraham tensor 
contains a phenomenological volume force.
\par
The macroscopic Maxwell equations of continuum electrodynamics
are the basis for deriving continuity equations for electric and
magnetic fields in a dielectric.
For a dielectric with no free charges in a regime of negligible
absorption and dispersion, the Maxwell equations may be written as
\begin{subequations}
\label{EQa3.01}
\begin{equation}
\nabla\times{\bf E}=-\frac{\partial{\bf B}}{\partial (ct)}
\label{EQa3.01a}
\end{equation}
\begin{equation}
\nabla\times{\bf B}=\frac{\partial n^2{\bf E}}{\partial (ct)} 
\label{EQa3.01b}
\end{equation}
\begin{equation}
\nabla\cdot{\bf B}=0
\label{EQa3.01c}
\end{equation}
\begin{equation}
\nabla\cdot \frac{n^2}{c^2}{\bf E}=0
\label{EQa3.01d}
\end{equation}
\end{subequations}
in Heaviside--Lorentz units.
The electric and magnetic fields can be defined in terms of the vector
potential ${\bf A}$ as
\begin{subequations}
\label{EQa3.02}
\begin{equation}
{\bf E}=-\frac{\partial {\bf A}}{\partial (ct)}
\label{EQa3.02a}
\end{equation}
\begin{equation}
{\bf B}= \nabla\times {\bf A}
\label{EQa3.02b}
\end{equation}
\end{subequations}
for transverse fields in the Coulomb gauge.
\par
The macroscopic Maxwell equations are the axioms of classical
continuum electrodynamics.
Poynting's theorem,
\begin{equation}
\frac{\partial} {\partial (ct)}
\left [ \frac{1}{2}\left ( n^2{\bf E}^2+{\bf B}^2  \right ) \right ]
+\nabla\cdot ( {\bf E}\times{\bf B}) = 0,
\label{EQa3.03}
\end{equation}
can be derived by multiplying the Faraday law (\ref{EQa3.01a}) by
${\bf B}$ and adding it to the Maxwell--Amp\`ere law (\ref{EQa3.01b})
multiplied by ${\bf E}$.
Poynting's theorem can also be derived by substituting the Maxwell
equations into the temporal derivative of the energy density
\begin{equation}
\rho_e=(1/2)(n^2{\bf E}^2+{\bf B}^2)
\label{EQa3.04}
\end{equation}
using a vector triple-product identity.
Poynting's theorem is a continuity equation for the Poynting
energy-flux vector
\begin{equation}
{\bf S}_P=c ({\bf E}\times{\bf B}) = ({s}^1_P,{s}^2_P,{s}^3_P).
\label{EQa3.05}
\end{equation}
The theorem
$$
\frac{\partial}{\partial (ct)} \left (n^2{\bf E}\times{\bf B}\right )=
-{\bf B}\times(\nabla\times{\bf B}) +{\bf B}(\nabla\cdot{\bf B})
$$
\begin{equation}
-n^2{\bf E}\times(\nabla\times{\bf E})+{\bf E}(\nabla\cdot n^2 {\bf E})
\label{EQa3.07}
\end{equation}
is derived, in a manner similar to Poynting's theorem, by substituting 
Maxwell's equations into the temporal derivative of the Minkowski
momentum density, ${\bf g}_M=(n^2/c)({\bf E}\times{\bf B})$.
The right-hand side of Eq.\ (\ref{EQa3.07}) can be recast, approximately,
as the negative of the divergence of the Maxwell stress
tensor \cite{BIJackson}
with components
\begin{equation}
W^{ij}
=\left [-n^2E_{i}E_{j}-B_{i}B_{j}+
\frac{1}{2}\left ( n^2{\bf E}\cdot{\bf E}+
{\bf B}\cdot{\bf B}\right )\delta_{ij}\right ],
\label{EQa3.08}
\end{equation}
where terms involving the gradient of $n^2$ have been neglected.
Then the temporal derivative of the Minkowski momentum density,
Eq.\ (\ref{EQa3.07}), can be expressed using the vector divergence
operator, $\bnab\;\cdot$, as
\begin{equation}
\frac{\partial}{\partial (ct)}(n^2{\bf E}\times{\bf B})
+\bnab \cdot {\bf W}=0.
\label{EQa3.09}
\end{equation}
The array
\begin{equation}
{T}_M^{\alpha\beta}=
\left [
\begin{matrix}
\rho_e
&{s}^1_P/c
&{s}^2_P/c
&{s}^3_P/c
\cr
c{g}_M^1
&W^{11}
&W^{12}
&W^{13}
\cr
c{g}_M^2
&W^{21}
&W^{22}
&W^{23}
\cr
c{g}_M^3
&W^{31}
&W^{32}
&W^{33}
\cr
\end{matrix}
\right ],
\label{EQa3.10}
\end{equation}
known as the Minkowski energy--momentum tensor,
is constructed from continuity equations (\ref{EQa3.03}) and
(\ref{EQa3.09}).
Using the summation convention, we can write Poynting's theorem
in Eq.\ (\ref{EQa3.03}) as
\begin{equation}
\partial_{0}T_M^{00}+\partial_{j}T_M^{0j}=0
\label{EQa3.11}
\end{equation}
for continuity of the energy flux and
Eq.\ (\ref{EQa3.09}) as
\begin{equation}
\partial_{0}T_M^{i0}+\partial_{j}T_M^{ij}=0
\label{EQa3.12}
\end{equation}
for continuity of the momentum flux.
Then, each row of $T_M$ corresponds to a four-divergence
\begin{equation}
\partial_{\beta}T_M^{\alpha\beta}=0.
\label{EQa3.13}
\end{equation}
Because the array $T_M^{\alpha\beta}$ in Eq.\ (\ref{EQa3.10}) is not
symmetric, Abraham proposed the energy--momentum tensor
\begin{equation}
{T}_A^{\alpha\beta}=
\left [
\begin{matrix}
\rho_e
&{s}^1_P/c
&{s}^2_P/c
&{s}^3_P/c
\cr
c{g}_A^1
&W^{11}
&W^{12}
&W^{13}
\cr
c{g}_A^2
&W^{21}
&W^{22}
&W^{23}
\cr
c{g}_A^3
&W^{31}
&W^{32}
&W^{33}
\cr
\end{matrix}
\right ],
\label{EQa3.14}
\end{equation}
where ${\bf g}_A$ is given by Eq.\ (\ref{EQa1.02}).
The four divergence of this tensor is
\begin{equation}
\partial_{\beta}T_A^{\alpha\beta}=-f^{\beta},
\label{EQa3.15}
\end{equation}
where $f^{\beta}$ is the Abraham force.
The Cartesian components of the Abraham force
\begin{equation}
{\bf f}=
\frac{\partial}{\partial (ct)}\left ((n^2-1){\bf E}\times{\bf B}\right )
\label{EQa3.16}
\end{equation}
are obtained by substituting the Abraham momentum density into the
continuity equation (\ref{EQa3.09}) and the time-like coordinate 
component is $f^0=0$.
\par
It has been widely reported in the literature that neither ${T}_M$
nor ${T}_A$ is the total energy--momentum tensor \cite{BIPfeifer}.
Instead, they are to be considered two of many arbitrary forms of the 
electromagnetic part, ${T}_{\rm fld}$, of a total energy momentum
tensor
\begin{equation}
{T}={T}_{\rm fld} +{T}_{\rm matl}
\label{EQa3.17}
\end{equation}
composed of energy--momentum tensors for the field and material
subspaces,
however those subspaces are defined.
\par
\section{Total Momentum}
\par
A well-defined quantity for the total momentum is only derivable
from the macroscopic Maxwell equations by imposing an additional
condition: The total momentum of an isolated system must be
constant in time.
In this section, we identify the unique total momentum using this
constraint.
We consider the case of a quasi-monochromatic electromagnetic field,
in the plane-wave limit, entering a linear medium from vacuum at
normal incidence.
The medium is taken to be a simple linear dielectric in the regime of
negligible dispersion and negligible absorption.
In the absence of reflection, there is no momentum given to the
dielectric slab and it remains stationary.
We adopt this case of a stationary dielectric in which reflections
can be neglected by assuming that an antireflection coating has been
applied to the dielectric or that the refractive index of the dielectric
is only slightly greater than unity.
\par
Gordon \cite{BIGord} used a microscopic model of the dielectric as a 
vapor of weakly polarizable atoms and derived the material momentum 
as the continuum average of the mechanical momentum of the atoms.
Assuming a rarefied vapor of atoms in order for reflections to be
negligible, Gordon obtained the total momentum
\begin{equation}
{\bf G}_{\rm G}=\int_V dv \, {\bf g}_G
=\int_V dv \frac{n}{c}({\bf E}\times{\bf B})
\label{EQa4.01}
\end{equation}
by adding the material momentum to the Abraham momentum.
Here the integration is over all three-dimensional space.
In continuum electrodynamics, the electrodynamic properties of a
material are characterized only by a macroscopic refractive index.
Therefore, the microscopic origin of the Gordon momentum (\ref{EQa4.01})
is of no consequence in the formalism of continuum electrodynamics.
Following Gordon, we will demonstrate that the momentum (\ref{EQa4.01})
is invariant in time and, because it is conserved, can be identified
as the total momentum.
Because the Gordon momentum depends on the material only through the
refractive index, it is the unique total momentum for all cases in which
the medium behaves, to a good approximation, as a negligibly reflecting
linear dielectric with refractive index $n$.
\par
Propagation of a field in a linear medium is governed by the wave
equation,
\begin{equation}
\nabla^2{\bf A}
-\frac{n^2}{c^2}\frac{\partial^2 {\bf A}}{\partial t^2}=0,
\label{EQa4.02}
\end{equation}
written in terms of the vector potential ${\bf A}$, where
${\bf B}=\nabla\times {\bf A}$.
For quasi-monochromatic plane waves, it is convenient to write the
vector potential in terms of a slowly varying envelope function,
$A(z,t)$, a rapidly varying carrier, and a unit vector,
${\bf e}_{\bf k}$, in the direction of propagation as
${\bf A}=A(z,t)e^{-i(\omega_0 t - k z)} {\bf e}_{\bf k}$.
For a plane-wave entering a dielectric at normal incidence,
reflections are negligible if $\delta n=n-1$ is small.
In this limit, there is no momentum given to the bulk material, which
remains stationary, and we can apply the Fresnel relation
\begin{equation}
\frac{{A}_t}{{A}_i}=\frac{2}{n+1}
=\left (1+\frac{\delta n}{2}\right )^{-1},
\label{EQa4.03}
\end{equation}
where $A_i$ is the incident amplitude and $A_t$ is the transmitted
amplitude.
Comparing Eq.\ (\ref{EQa4.03}) with a series expansion of $1/\sqrt{n}$
in the limit of small $\delta n$, we find that the vector
potential amplitude inside the dielectric, $A_t$, is reduced by
a factor of $\sqrt{n}$ from the incident amplitude $A_i$.
\par
For continuous plane waves, and approximately for slowly varying waves,
the relation between the amplitudes of the fields simplifies to
\begin{equation}
|{\bf B}|=n|{\bf E}|= \frac{n\omega}{c} |{\bf A}|.
\label{EQa4.04}
\end{equation}
Then, the electromagnetic energy density,
$\rho_e=(1/2)(n^2{\bf E}^2+{\bf B}^2)$,
can be written as
\begin{equation}
\rho_e = \frac{n^2\omega^2}{c^2}|{\bf A}|^2.
\label{EQa4.05}
\end{equation}
Applying the Fresnel amplitudes to relate the fields inside and outside
the medium, $A_t=A_i/\sqrt{n}$, we find that the energy density inside
the material is a factor of $n$ larger than the energy density of the
same field in the vacuum.
The region occupied by the field in the material is compressed spatially
by a factor of $n$ due to the reduced speed of light within the medium,
such that the total energy 
\begin{equation}
U=\int_V dv \rho_e=\int_V dv \frac{n^2\omega^2}{c^2}|{\bf A}|^2
\label{EQa4.06}
\end{equation}
is conserved.
Numerical solutions of the wave equation for a field entering a linear
material through a gradient-index anti-reflection coating indicate that
the field in the material is a factor of $\sqrt{n}$ smaller and a factor
of $n$ narrower than the field in the vacuum, independent of the
magnitude of $n$, as long as reflections are suppressed \cite{BINum}.
\par
Having demonstrated the conservation properties of the electromagnetic
energy, we demonstrate the conservation properties of the
Gordon electromagnetic momentum by a similar procedure.
The Gordon momentum \cite{BIGord} is obtained by integrating the
momentum density (\ref{EQa1.03}) over all three-dimensional space.
Comparing the Gordon momentum, expressed in terms of the envelope
functions
\begin{equation}
{\bf G}_{\rm G}=\int_V dv
\frac{n}{c}\left ({\bf E}\times{\bf B} \right )
=\int_V dv \frac{n^2\omega^2}{c^3}|{\bf A}|^2 {\bf e}_k,
\label{EQa4.08}
\end{equation}
with the total energy in Eq.\ (\ref{EQa4.06}), we see that conservation of
total energy implies conservation of the Gordon momentum.
The total momentum of a closed system is unique and 
the Gordon form of total momentum is conserved.
Therefore, the Gordon momentum can be identified as the total momentum
of the thermodynamically closed system.
Because neither the Minkowski momentum nor the Abraham momentum
is the total momentum, neither is conserved in a thermodynamically
closed system.
\par
We point out that, in the macroscopic limit in which the dielectric is
described by a refractive index $n$, the question of what portion of
the energy given by Eq.\ (\ref{EQa4.06}) resides in the field or
dielectric is improperly posed.
Comparing Eqs.\ (\ref{EQa4.06}) and (\ref{EQa4.08}), we see that the
same holds true for the apportionment of the momentum into field and
dielectric components.
\par
\section{Total Energy--Momentum Tensor}
\par
In the previous section, we identified the unique total momentum for the
system of an electromagnetic field in a dielectric for the case of 
negligible reflections.
In this section, we construct the corresponding total energy--momentum
tensor from continuity equations for the energy and momentum.
\par
A continuity equation is a differential form of a conservation law
applied to an element of volume in a continuous flow.
In Sec. II we showed that the energy--momentum tensor for dust is
constructed from continuity equations in which the differential
operators act on the densities of conserved quantities.
The operand of the time derivative in the continuity
Eq.\ (\ref{EQa3.09}) is the Minkowski momentum density.
Because the volume integral of the Minkowski momentum density is not
a conserved vector quantity, we do not consider Eq.\ (\ref{EQa3.09})
to be a suitable continuity equation with which to
construct an energy--momentum tensor.
Instead, we write Eq.\ (\ref{EQa3.09}) as
\begin{equation}
\frac{n}{c}\frac{\partial}{\partial t}c{\bf g}_{\rm G}
+\bnab \cdot {\bf W}=0,
\label{EQa5.01}
\end{equation}
in terms of the Gordon momentum density in Eq.\ (\ref{EQa1.03})
whose volume integral is a conserved vector quantity.
Equation (\ref{EQa5.01}) provides three continuity equations for our
energy--momentum tensor.
The additional continuity equation is obtained by writing Poynting's
theorem in
(\ref{EQa3.03}) as
\begin{equation}
\frac{n}{c}\frac{\partial \rho_e} {\partial t}
+\nabla\cdot [n ({\bf E}\times{\bf B})]
=\frac{\nabla n}{n} \cdot n( {\bf E}\times{\bf B})
\label{EQa5.02}
\end{equation}
using the densities of conserved quantities.
Again, we are considering the case of a closed system in which there
are no reflections.
As this is assumed to be accomplished by a gradient-index antireflection
coating, we can drop the term containing $\nabla n$ and write Poynting's
theorem in Eq.\ (\ref{EQa5.02}) as
\begin{equation}
\frac{n}{c}\frac{\partial \rho_e} {\partial t}
+\nabla\cdot [n( {\bf E}\times{\bf B})]=0.
\label{EQa5.03}
\end{equation}
\par
In Maxwell's equations, expressed in terms of 3-vectors, time is not a
coordinate.
We make a simple change of time variable to $\tau=t/n$ and write
the continuity theorems (\ref{EQa5.01}) and (\ref{EQa5.03}) as
\begin{subequations}
\label{EQa5.04}
\begin{equation}
\frac{1}{c}\frac{\partial}{\partial \tau}c{\bf g}_{\rm G}
+\bnab \cdot {\bf W}=0
\label{EQa5.04a}
\end{equation}
\begin{equation}
\frac{1}{c}\frac{\partial \rho_e} {\partial \tau}
+\nabla\cdot c{\bf g}_{G}=0,
\label{EQa5.04b}
\end{equation}
\end{subequations}
where the Gordon momentum density ${\bf g}_G$ is given in
Eq.\ (\ref{EQa1.03}).
However when writing Maxwell's equations as tensor equations, time is
one of the four space--time coordinates and we define the
time-like coordinate
\begin{equation}
\bar x^0=c\tau=\frac{ct}{n}.
\label{EQa5.05}
\end{equation}
Then the four scalar continuity equations, Eqs.\ (\ref{EQa5.04a}) and
(\ref{EQa5.04b}), can be written concisely as a single equation, as in
Section 2,
defining the operator
\begin{equation}
\bar\partial_{\alpha}
=\left ( \frac{\partial}{\partial \bar x^0},
\partial_x, \partial_y,\partial_z \right )
\label{EQa5.06}
\end{equation}
and an array
\begin{equation}
T^{\alpha\beta}= \left [
\begin{matrix}
\rho_e
&c{g}_{\rm G}^1
&c{g}_{\rm G}^2
&c{g}_{\rm G}^3
\cr
c{g}_{\rm G}^1
&W^{11}
&W^{12}
&W^{13}
\cr
c{g}_{\rm G}^2
&W^{21}
&W^{22}
&W^{23}
\cr
c{g}_{\rm G}^3
&W^{31}
&W^{32}
&W^{33}
\cr
\end{matrix}
\right ] ,
\label{EQa5.07}
\end{equation}
such that
\begin{equation}
\bar\partial_{\beta}T^{\alpha\beta}=0.
\label{EQa5.08}
\end{equation}
\par
The array that appears in Eq.\ (\ref{EQa5.07}) has a number of
notable properties.
The array is symmetric
\begin{equation}
T^{\alpha\beta}=T^{\beta\alpha}
\label{EQa5.09}
\end{equation}
and has a vanishing trace
\begin{equation}
T^{\alpha}_{\alpha}= 0.
\label{EQa5.10}
\end{equation}
The operator defined in Eq.\ (\ref{EQa5.06}) applied to the rows
of the array in Eq.\ (\ref{EQa5.07}) generates continuity laws for
demonstrably conserved electromagnetic energy and momentum properties.
Similarly, the operator in Eq.\ (\ref{EQa5.06}) applied to the columns,
\begin{equation}
\bar\partial_{\alpha} T^{\alpha\beta}=0,
\label{EQa5.11}
\end{equation}
generates the same continuity equations as a consequence of the
symmetry of the array.
These are the properties that we associate with an energy--momentum
tensor.
Ravndal \cite{BIRav} arrives at the same energy--momentum tensor
as in Eq.\ (\ref{EQa5.07}) using symmetry arguments, but interprets the 
continuity equations, Eqs.\ (\ref{EQa5.08}), in the context of the
Minkowski momentum, which we have shown is not conserved.
\par
\section{Concluding Remarks}
\par
For many years, the Abraham--Minkowski controversy has been resolved by
postulating a total energy--momentum tensor that is comprised of
separate field and matter tensors --- an approach that involves assumptions
about the behavior of matter in the presence of an electromagnetic
field.
In continuum electrodynamics the interaction of the field and matter is
described in terms of a single macroscopic parameter, the refractive
index $n$.
We showed that the Gordon momentum is the total momentum of a
thermodynamically closed system consisting of a quasimonochromatic
field and negligibly reflecting linear dielectric in the continuum.
We derived continuity equations from the Maxwell equations and used
a time variable transformation to write the continuity equations in
terms of densities of conserved energy and Gordon momentum quantities.
When written in four-dimensional tensor form with time-like
coordinate $ct/n$, the
continuity equations are obtained from the four-divergence of a
traceless, diagonally symmetric energy--momentum tensor.
\par
In summary, for the case of an electromagnetic field and negligibly
reflecting dielectric, we constructed the total energy--momentum
tensor in Eq.\ (\ref{EQa5.07}) from continuity equations that were
derived from the macroscopic Maxwell's equations.
It is interesting to note that a time coordinate transformation was
required in order to write the continuity equations as the
four-divergence of the symmetric energy--momentum tensor.
\vfill
\par

\end{document}